# Розрахунок фаз *np*- розсіяння до $T_{lab}$=3 ГеВ для потенціалів Аргоннської групи за методом фазових функцій

## В.І. Жаба

*Ужгородський національний університет, кафедра теоретичної фізики,*
*вул. Волошина, 54, Ужгород, 88000, Україна*

Для обрахунку фаз одноканального нуклон-нуклонного розсіяння розглянуто відомий метод фазових функцій. За допомогою методу фазових функцій чисельно отримано фазові зсуви *np*- розсіяння для $^1S_0$-, $^1P_1$-, $^3P_0$-, $^3P_1$-, $^1D_2$-, $^3D_2$-, $^1F_3$-, $^3F_3$-, $^1G_4$-, $^3G_4$- станів. Розрахунки проведено для сучасних реалістичних нуклон-нуклонних потенціалів Аргоннської групи (Av18, Av14sscc, Av8', Av8'sscc, Av6', Av4', Av2'). Чисельно розраховані фазові зсуви добре узгоджуються з результатами, отриманими іншими методами. По розрахованим фазовим зсувам обчислено повний переріз *np*- розсіяння.

**Ключові слова:** розсіяння, метод, фаза, нуклон, стан.

# Calculation of phases of *np*- scattering up to $T_{lab}$=3 GeV for potentials Argonne group on the phase-function method

## V.I. Zhaba

*Uzhgorod National University, Department of Theoretical Physics,*
*54 Voloshyna St., Uzhgorod, UA-88000, Ukraine*

For calculation of the single-channel nucleon-nucleon scattering a phase-functions method has been proposed. Using a phase-functions method the following phase shifts of a *np*- scattering numerically for $^1S_0$-, $^1P_1$-, $^3P_0$-, $^3P_1$-, $^1D_2$-, $^3D_2$-, $^1F_3$-, $^3F_3$-, $^1G_4$-, $^3G_4$- states are calculated. The calculations has been performed using realistic nucleon-nucleon potentials Argonne group (Av18, Av14sscc, Av8', Av8'sscc, Av6', Av4', Av2'). Obtained phase shifts are in good agreement with the results obtained in the framework of other methods. Using the obtained phase shifts we have calculated the full cross-section *np*- scattering.

# Вступ

Із експериментально спостережуваних величин перерізу розсіяння та енергій переходів отримують у першу чергу інформацію про фази та амплітуди розсіяння, ніж про хвильові функції, що є основним об'єктом дослідження при стандартному підході. Іншими словами, в експерименті спостерігаються не самі хвильові функції, а їх зміни, викликані у результаті взаємодії [1]. Тому представляє інтерес розглядати такі рівняння, що безпосередньо пов'язують фази й амплітуди розсіяння з потенціалом, не знаходячи при цьому хвильові функції.

Точний розв'язок задачі розсіяння із метою обчислення фаз розсіяння можливе тільки для окремих феноменологічних потенціалів. Коли використовуються реалістичні потенціали, то фази розсіяння обчислюються наближено. Це пов'язано з використанням фізичних апроксимацій або з чисельним розрахунком. Вплив вибору чисельного алгоритму на рішення задачі розсіяння вказано у роботі [2].

В останні 10 років зріс інтерес до нуклон-нуклонного розсіяння у рамках кіральної теорії збурень [3,4], в послідовному теоретико-польовому підході [5], для парціального хвильового аналізу нижче порогу утворення піона [6]. Також отримуються фази розсіяння через суперсиметрії і факторизації [7], N/D метод обчислення парціальних хвиль для еластичного *NN*- розсіяння [8] або перенормування *NN*- взаємодії для кірального потенціалу двопіонного обміну [9].

До методів розв'язування рівняння Шредінгера з метою отримання фаз розсіяння належать: метод послідовних наближень, борнівське наближення, метод фазових функцій та інші. Метод фазових функцій виявився досить зручним при розв'язуванні багатьох конкретних задач атомної і ядерної фізики.

Основною і головною перевагою методу фазових функцій (МФФ) при застосуванні до задач нуклон-нуклонного розсіяння є та, що МФФ дозволяє отримати фази розсіяння, не знаходячи при цьому хвильові функції як розв'язки рівняння Шредінгера. Завдяки фазовому рівнянню наявний безпосередній зв'язок між фазою розсіяння і потенціалом взаємодії.

У минулому столітті ММФ частіше використовувався. Але все ж таки у дослідників залишається певний інтерес до його застосування для розрахунків. Наприклад, метод квазілінеаризації [10] з МФФ дає чудові результати, коли застосовується до обчислення основних і збуджених зв'язаних станів енергій і хвильових функцій для різних потенціалів. У роботі [11] запропоновано узагальнення МФФ до проблем акустичної хвилі, що розсіюється на безперервній середній неоднорідності.

Дана робота присвячена розрахунку фазових зсувів *np*- розсіяння у різних спінових станах для сучасних реалістичних феноменологічних нуклон-нуклонних потенціалів Аргоннської групи [12,13] за допомогою методу фазових функцій (МФФ).

## Метод фазових функцій

Математично МФФ - це особливий спосіб розв'язування радіального рівняння Шредінгера

$$u''_l(r) + \left(k^2 - \frac{l(l+1)}{r^2} - U(r)\right) u_l(r) = 0, \tag{1}$$

яке є лінійним диференціальним рівнянням другого порядку. В (1) величина $U(r) = \frac{2m}{\hbar^2} V(r)$ - це перенормований потенціал взаємодії, *m* - приведена маса. МФФ досить зручний для отримання фаз розсіяння, оскільки по цьому методу не потрібно спочатку обчислювати в широкій області радіальні хвильові функції задачі розсіяння і потім по їх асимптотикам знаходити ці фази.

Стандартний спосіб обчислення фаз розсіяння - це розв'язок рівняння Шредінгера (1) з асимптотичною граничною умовою. МФФ - це перехід від рівняння Шредінгера до рівняння для фазової функції. Для цього роблять заміну [1,14]:

$$u_l(r) = A_l(r) \left[ \cos\delta_l(r) \cdot j_l(kr) - \sin\delta_l(r) \cdot n_l(kr) \right]. \tag{2}$$

Введені дві нові функції $\delta_l(r)$ і $A_l(r)$ мають зміст відповідних фаз розсіяння і констант нормування (амплітуд) хвильових функцій для розсіяння на визначеній послідовності обрізаних потенціалів. $\delta_l(r)$ і $A_l(r)$ називаються відповідно їх фізичному змісту фазовою й амплітудною функцією. Термін "фазова функція" вперше був використаний у роботі Морзе і Алліса [15]. Рівняннями для фазової й амплітудної функцій з початковими умовами є:

$$\delta'_l(r) = -\frac{1}{k} U(r) \left[ \cos\delta_l(r) \cdot j_l(kr) - \sin\delta_l(r) \cdot n_l(kr) \right]^2, \quad \delta_l(0) = 0; \tag{3}$$

$$A'_l(r) = -\frac{1}{k} A_l(r) U(r) \left[ \cos\delta_l(r) \cdot j_l(kr) - \sin\delta_l(r) \cdot n_l(kr) \right] \times \\ \times \left[ \sin\delta_l(r) \cdot j_l(kr) + \cos\delta_l(r) \cdot n_l(kr) \right], \quad A_l(0) = 1. \tag{4}$$

Фазове рівняння (3) було вперше отримано Друкарєвим, а потім незалежно у роботах Бергмана, Колоджеро і Зимека. Частинний випадок рівняння (4) при *l=0* був використаний Морзе і Аллісом при дослідженні задачі *S*- розсіяння повільних електронів на атомах [15].

Відмітимо переваги підходу МФФ для обчислення фаз у порівнянні зі стандартним методом, заснованим на розгляді рівняння Шредінгера для хвильової функції [1]. Той факт, що фазове рівняння - першого порядку (хоч і нелінійне), спрощує програмування і обчислення на ЕОМ. Крім того, при цьому зменшується кількість операцій і час розрахунку. Не осцилюючий, а більш монотонний характер поведінки фазової функції дозволяє проводити розрахунки з більшою точністю і полегшує оцінку похибок результатів. Ще однією важливою перевагою даного методу є можливість побудови в його рамках нових алгоритмів обчислення не тільки фазових зсувів, але й парціальних і повних амплітуд розсіяння, елементів $S$- матриці, довжин розсіяння, ефективних радіусів та інших параметрів розсіяння.

## Потенціали Аргоннської групи

До сучасних нуклон-нуклонних потенціалів взаємодії відносяться наступні: залежний від заряду потенціал CD-Bonn [16], потенціал Moscow [17], кіральний потенціал Айдахо [18], релятивістська оптична модель [19] на базі Московського потенціалу, бразильський релятивістський потенціал двопіонного обміну $O(q^4)$ [20], локальний нуклон-нуклонний потенціал, розширений з точки зору ортогональних проекторів [21], потенціали Неймегенської групою [22].

Чому саме вибрано провести розрахунок фазових зсувів з використанням потенціалів Аргоннської групи? Параметри потенційних моделей оптимізовані таким чином, що мінімізовано значення $\chi^2$ у прямій підгонці до даних. Для потенціалу Nijm92pp величина $\chi^2/N_{pp}$ становила 1,4. Наступне удосконалення потенціалу Nijm78 для $np$ даних дало модель Nijm93: $\chi^2/N_{pp}$=1,8 для 1787 $pp$ і $\chi^2/N_{nn}$=1,9 для 2514 $np$ даних, тобто $\chi^2/N_{data}$=1,87. Для потенціалів Nijm I і NijmII величина $\chi^2/N_{data}$=1,03. Оригінальний потенціал Рейда Reid68 був параметризований на основі фазового аналізу і отримав назву Reid93. Параметризація була проведена для 50 параметрів $A_{ij}$ і $B_{ij}$ потенціалу, причому $\chi^2/N_{data}$=1,03 [22].

Потенціал Argonne v18 [12] з 40 регульованими параметрами дає величину $\chi^2/N_{data}$=1,09 для 4301 $pp$ і $np$ даних в області енергій 0-350 МеВ. Для потенціалу CD-Bonn [16] величина $\chi^2/N_{data}$ становить 1,01 для 2932 $pp$ даних і 1,02 для 3058 $np$ даних. Такі потенціали, як Хамада-Джонсона-62, потенціали Єльської групи, Reid68, UrbanaV14 та ін. мають більші значення $\chi^2$, оскільки параметризовані на основі більш вузького енергетичного інтервалу.

Отже, потенціали Аргоннської групи є одними з тих реалістичних феноменологічних потенціалів, які найкраще описують міжнуклонну взаємодію.

При розрахунках фаз розсіяння потрібно враховувати особливості потенціалу Argonne v18, що складається з суми [12]

$$V = V^{EM} + V^{\pi} + V^{R}, \qquad (5)$$

де $V^{EM}$ – електромагнітна частина; $V^{\pi}$ – частина, що описує однопіонний обмін (OPE); $V^{R}$ – проміжкова і короткодіюча феноменологічна частина потенціалу. Слід зауважити, що для *pp*- розсіяння електромагнітна частина потенціалу включає одно- і дво- фотонні кулонівські терми, терм Дарвіна-Фолді, вакуумну поляризацію і взаємодію між магнітними моментами:

$$V^{EM}(pp) = V_{C1}(pp) + V_{C2} + V_{DF} + V_{VP} + V_{MM}(pp). \qquad (6)$$

Для *np*- системи електромагнітна частина потенціалу складається з терму Кулона, що відноситься до розподілу нейтронного заряду, і взаємодії між магнітними моментами:

$$V^{EM}(np) = V_{C1}(np) + V_{MM}(np). \qquad (7)$$

Для *nn*- системи нехтують кулонівською взаємодією між нейтронними формфакторами, і потенціал визначається тільки термом магнітного моменту. Крім цього, для розрахунку фаз розсіяння для $^1S_0$- стану для *pp*- системи враховують повну електромагнітну взаємодію і відносну електромагнітну хвилю, а для *nn*- і *np*- систем – тільки повну електромагнітну взаємодію.

Особливості інших потенціалів Аргоннської групи описано в роботі [13].

**Розрахунки фазових зсувів і обговорення результатів**

Методом фазових функцій чисельно розраховано фазові зсуви нуклон-нуклонного *np*-розсіяння для $^1S_0$-, $^1P_1$-, $^3P_0$-, $^3P_1$-, $^1D_2$-, $^3D_2$-, $^1F_3$-, $^3F_3$-, $^1G_4$-, $^3G_4$- станів. Маси нуклонів вибрано такими: $M_p$=938,27231 МеВ; $M_n$=939,56563 МеВ. Був вибраний чисельний метод розв'язання фазового рівняння (3) - метод Рунге-Кутта четвертого порядку [23]. Програмний код для чисельних розрахунків написаний на мові програмування Фортран. При оптимізованому виборі кроку чисельних розрахунків фазові зсуви отримувалися з точністю до 0,01. Фазові зсуви знаходились при виході фазової функції $\delta_l(r)$ на асимптотику при *r*>25 Фм. Значення фазових зсувів приведено на рис. 1-14. Фазові зсуви вказано у градусах. Чисельні розрахунки проведено для потенціалів Аргоннської групи (Av18, Av14sscc, Av8', Av8'sscc, Av6', Av4', Av2'). Діапазон енергій 1-3000 МеВ.

Для області енергій 1-400 МеВ наявне добре узгодження між фазовими зсувами отриманими на основі МФФ (результати даної роботи) і даними в інших роботах (для Av18 [12,24], Av14 [13], Av8', Av6', Av4' [24]). Розходження між результатами становить не більше двох відсотків.

Слід зауважити, що в останні 15 років зріс інтерес для знаходження фазових зсувів при великих енергіях. Нажаль, у літературних джерелах значно менше наявних розрахованих фазових зсувів при великих енергій для потенціалів Аргоннської групи. У роботі [25] приведені фазові зсуви для потенціалу Av18. Причому розрахунки проведено до 1000 МеВ лише для станів $^1S_0$-, $^3P_0$- і $^3F_3$. В [26] приведено результати розрахунків фазових зсувів до 1,6 ГеВ для потенціалів інверсії, що базуються на SM94, OSBEP, Av18 і Bonn-B. Крім цього згідно [27] потенціали Nijm-1, Nijm-2, Av18 та квантової інверсії Джел'фанд-Левітан-Марченко були розширені як *NN* оптичні моделі. Причому було враховано аналіз фазових зсувів по *Арндту та ін.* (SP00, FA00, WI00) від 300 МеВ до 3 ГеВ. В [28] наявні фазові зсуви тільки до 1000 МеВ для $^1F_3$- стану для потенціалів Nijm-1, Nijm-2, Reid93, Bonn і Av18, які були екстрапольовані для високих енергій.

Відмінність між отриманими фазовими зсувами по МФФ і даними [25-28] для потенціалу Av18 становить не більше 5 відсотків. Зрештою розраховані фазові зсуви для конкретної спінової конфігурації при великих енергіях (більше 350 МеВ) для потенціалів Аргоннської групи відрізняються між собою у більшості випадків.

Крім результатів у згаданих роботах, фазові зсуви розраховано до 1000 МеВ для нуклон-нуклонних моделей CD-Bonn і N3LO [25], а також для Arndt, OBEP, Bonn-B, Nijm-3 і Paris [29]. Для релятивістської оптичної моделі на базі Московського потенціалу [19] та потенціалів MYQ2, MYQ3, MY2, SP07 і Graz II [30] фазові зсуви отримано до 3 ГеВ, а для потенціалу Дірака [31] до 1,2-3,0 ГеВ і для потенціалу інверсії і Паризького потенціалу [32] до 1,2 ГеВ.

Якщо ж порівнювати отримані по МФФ фазові зсуви при великих енергіях для потенціалів Аргоннської групи з даними для вказаних інших потенціальних моделей, то очевидне розходження між ними. Звісно це пов'язано з особливостями структури нуклон-нуклонних потенціалів.

Незважаючи на очевидні поліпшення в описі даних при енергіях нижче 400 МеВ, теоретичні розрахунки досі показують деякі значні систематичні недоліки [32]. Зокрема, незрозуміла розбіжність спінових спостережуваних при передачах імпульсу нижче 1 Фм$^{-1}$. Походження таких розбіжностей з даними можна було б віднести до феноменологічного обмеження «голого» нуклон-нуклонного потенціалу особливо при більш високих енергіях, а також і до спрощень в моделі для *NN* ефективної взаємодії або до того факту, що оптична модель потенціалу тільки була розроблена до найнижчого порядку. Вище 350-400 МеВ розбіжності між потенціалами *NN* взаємодії стають ще більш очевидними. При цих енергіях більше 400 МеВ *NN* потенціали мають прикладне застосування. Тому оцінка теорії потребує більш точного опису «голої» взаємодії двох нуклонів.

Поряд з фазовими зсувами у задачах розсіяння приходиться мати справу з амплітудами розсіяння, елементами *S*- матриці і цілим рядом інших параметрів. По відомим фазам розсіяння отримують повну амплітуду, повний переріз і парціальну амплітуду розсіяння відповідно [1]

$$F(\theta) = \frac{1}{k}\sum_{l=0}^{\infty}(2l+1)e^{i\delta_l}\sin\delta_l P_l(\cos\theta), \qquad (8)$$

$$\sigma = \frac{4\pi}{k^2}\sum_{l=0}^{\infty}(2l+1)\sin^2\delta_l, \qquad (9)$$

$$f_l = \frac{1}{k}e^{i\delta_l}\sin\delta_l, \qquad (10)$$

де *P$_l$(cosθ)* - поліноми Лежандра, *θ* - полярний кут.

У роботі [33] вказаний повний переріз розсіяння, розрахований по фазових зсувах при енергіях 1-350 МеВ для потенціалів Неймегенської групи і потенціалу Av18. Використовуючи фазові зсуви при енергіях до 350 МеВ, результати розрахунків параметрів Волфштейна як функції кута ц.м. та $E_{LAB}$ наведені в [34] для PWA і Av18, а в [35] - для Reid93 та NijmII.

Результати розрахунків повного перерізу *np*- розсіяння (9) приведено на рис. 15. Чітко простежується відмінність отриманого повного перерізу в залежності від потенціалу Аргоннської групи.

Отже, результатами роботи є:

1. Вперше методом фазових функцій розраховано *np* фазові зсуви для відповідних спінових конфігурацій в області енергій від 1 МеВ до 3 ГеВ для семи нуклон-нуклонних потенціалів Аргоннської групи (Av18, Av14sscc, Av8', Av8'sscc, Av6', Av4', Av2').

2. Чисельно отримані фазові зсуви добре узгоджуються з результатами інших робіт для цих же потенціалів (відхилення становить не більше 5 відсотків).

3. По розрахованим фазовим зсувам за МФФ обчислено повний переріз *np*- розсіяння.

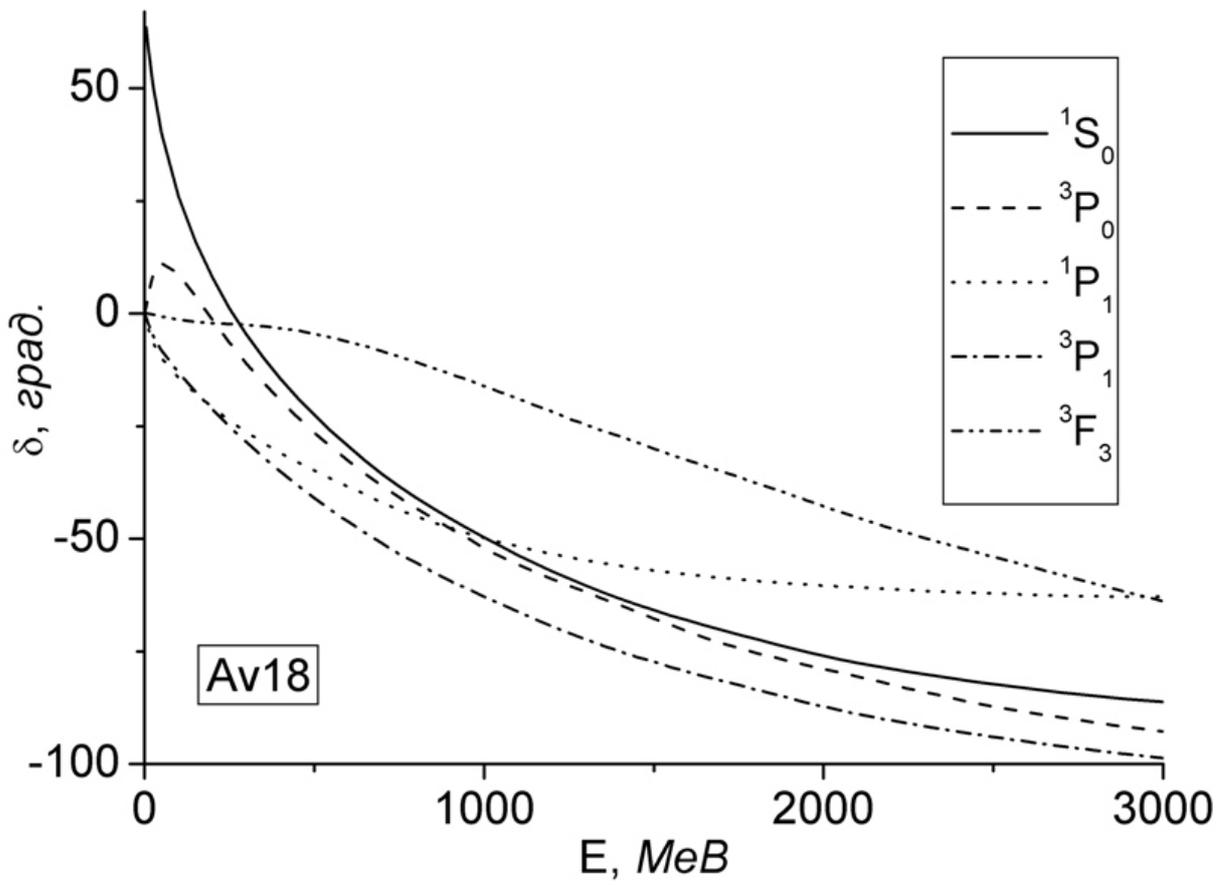

Рис. 1. Фазові зсуви *np*- розсіяння для потенціалу Av18

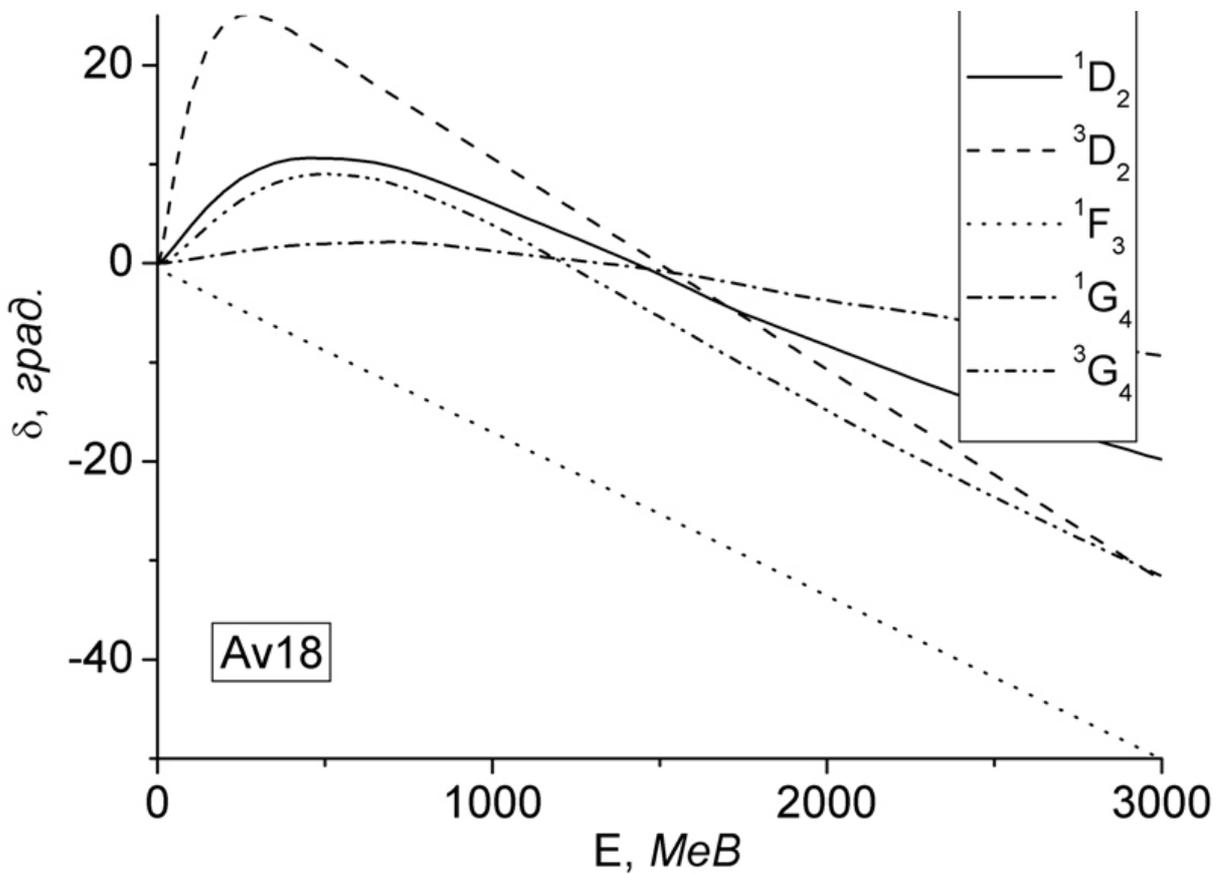

Рис. 2. Фазові зсуви *np*- розсіяння для потенціалу Av18

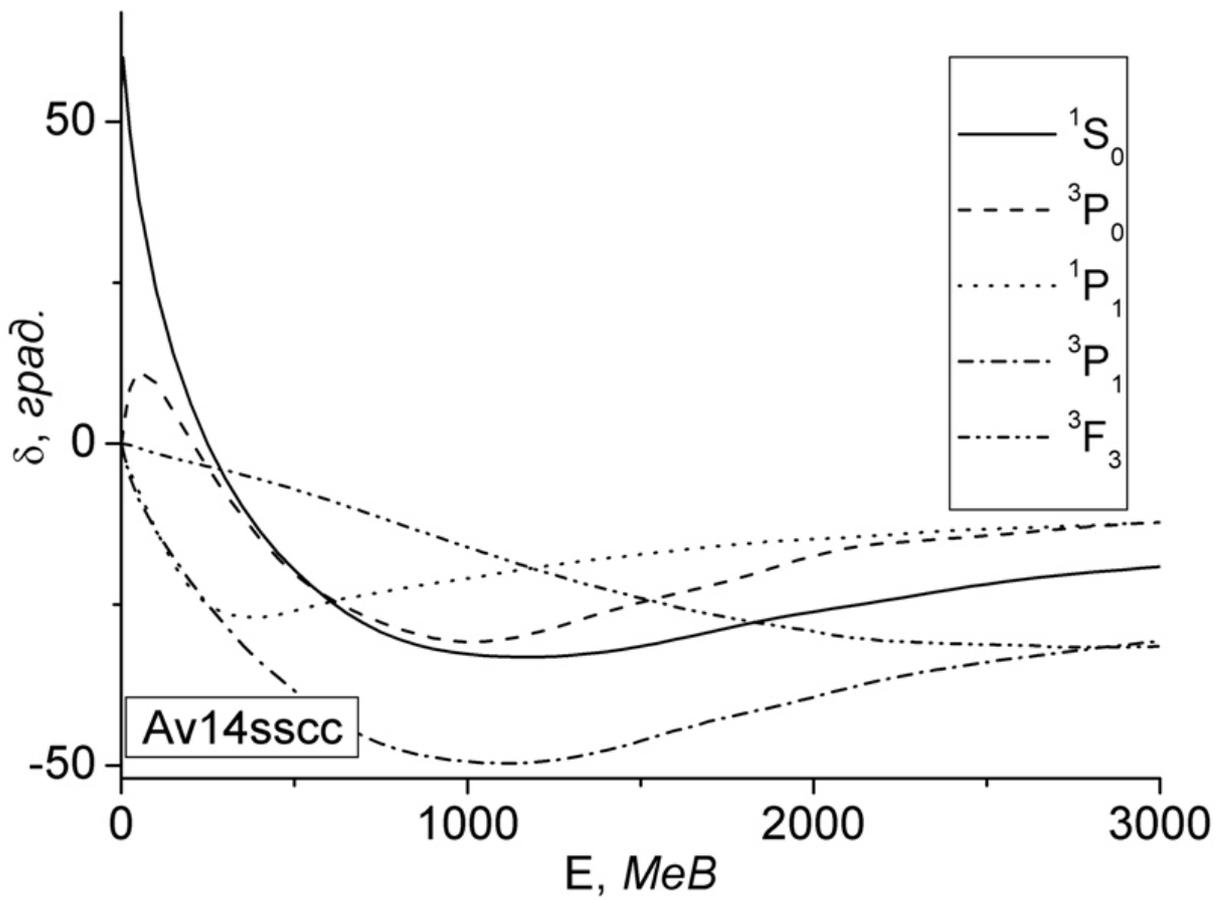

Рис. 3. Фазові зсуви *np*- розсіяння для потенціалу Av14sscc

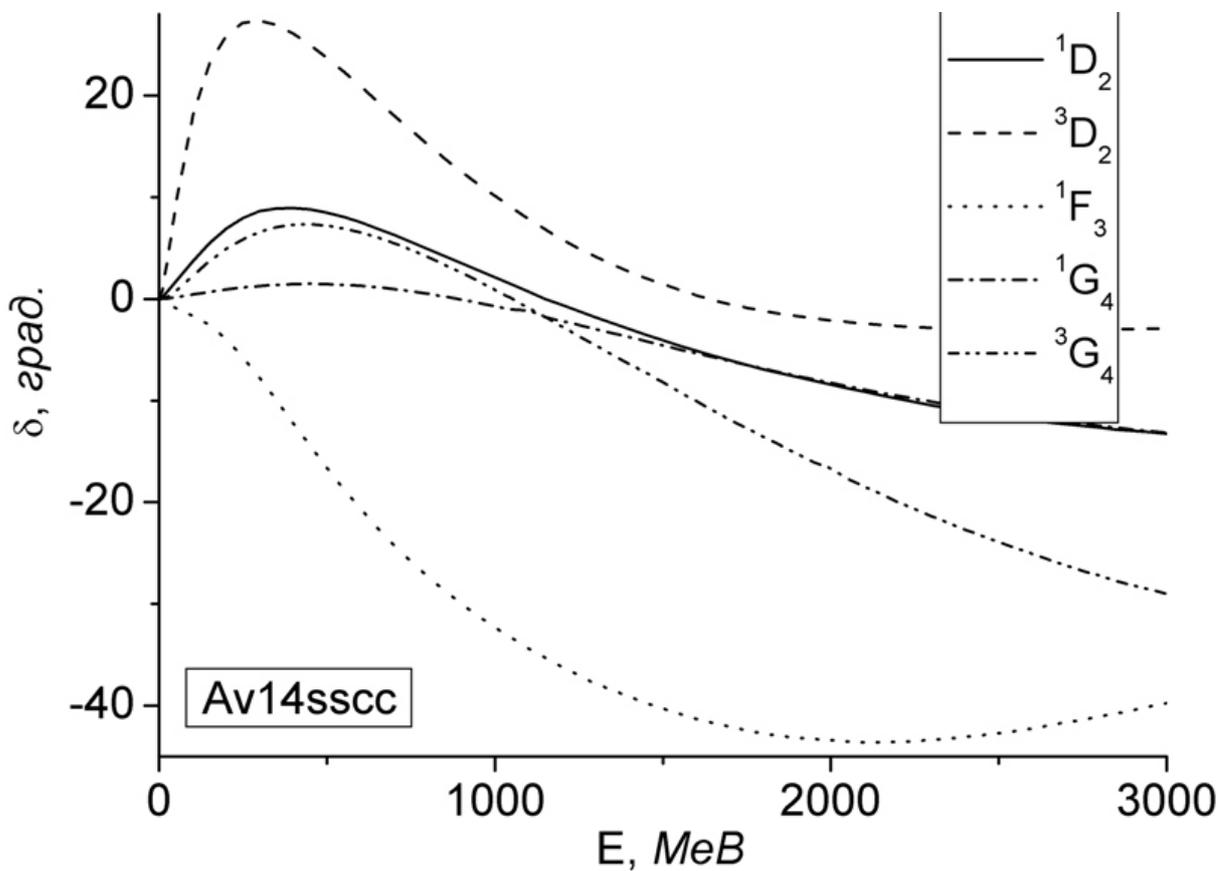

Рис. 4. Фазові зсуви *np*- розсіяння для потенціалу Av14sscc

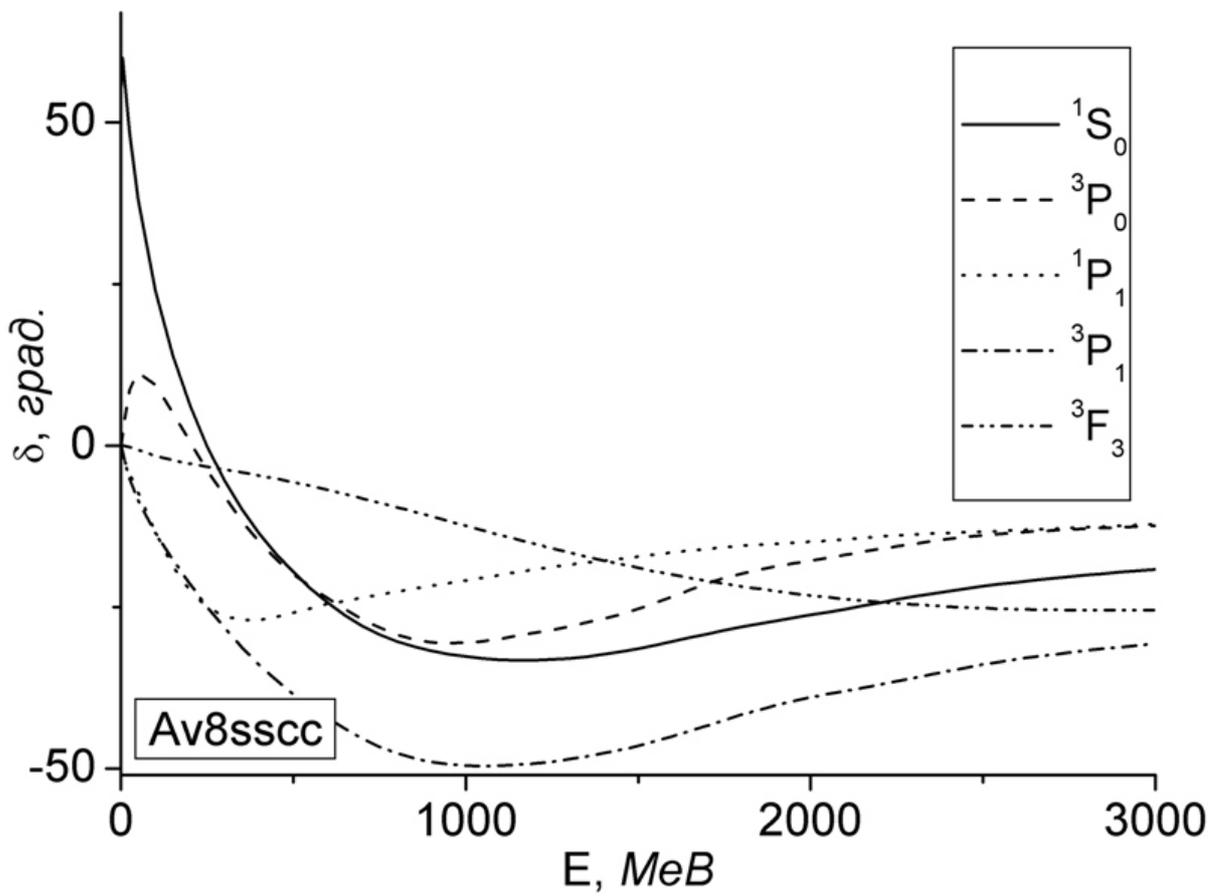

Рис. 5. Фазові зсуви *np*- розсіяння для потенціалу Av8'

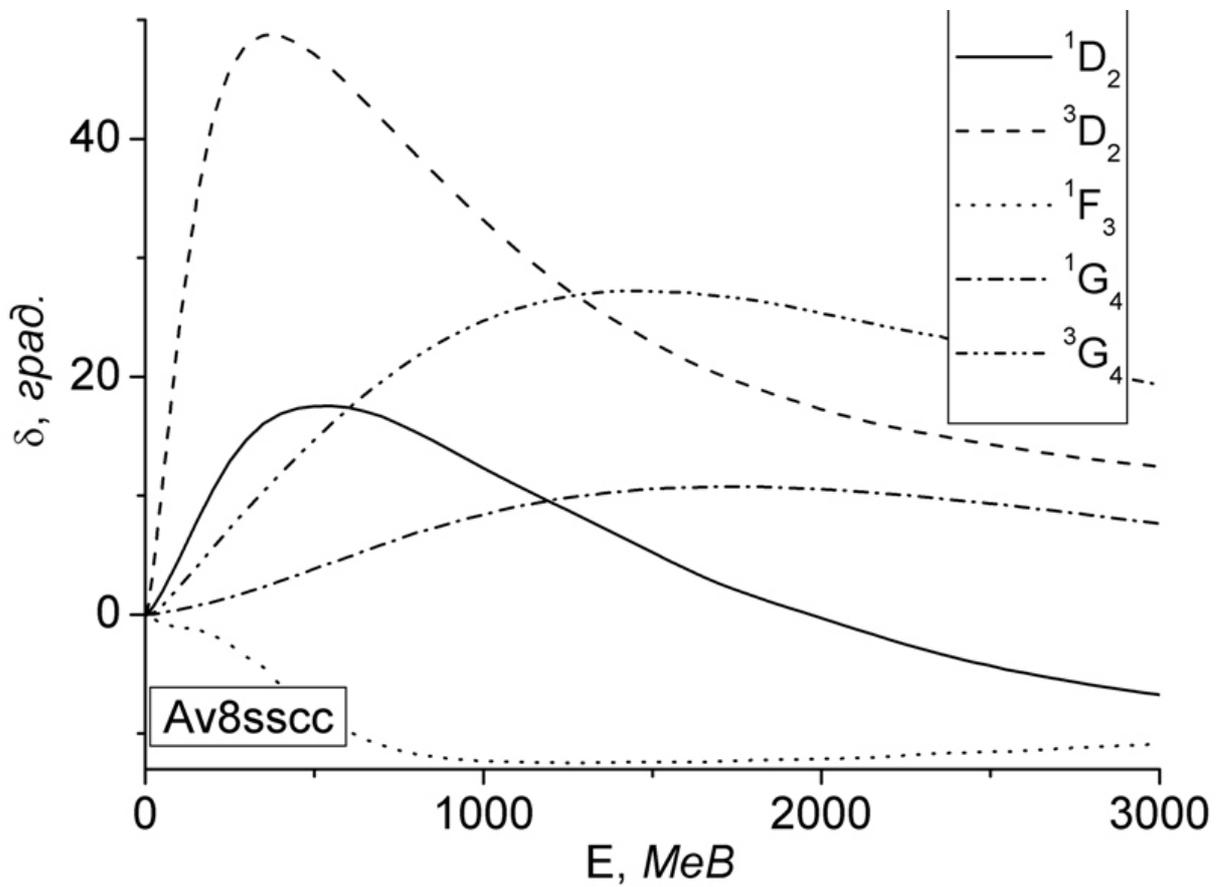

Рис. 6. Фазові зсуви *np*- розсіяння для потенціалу Av8'

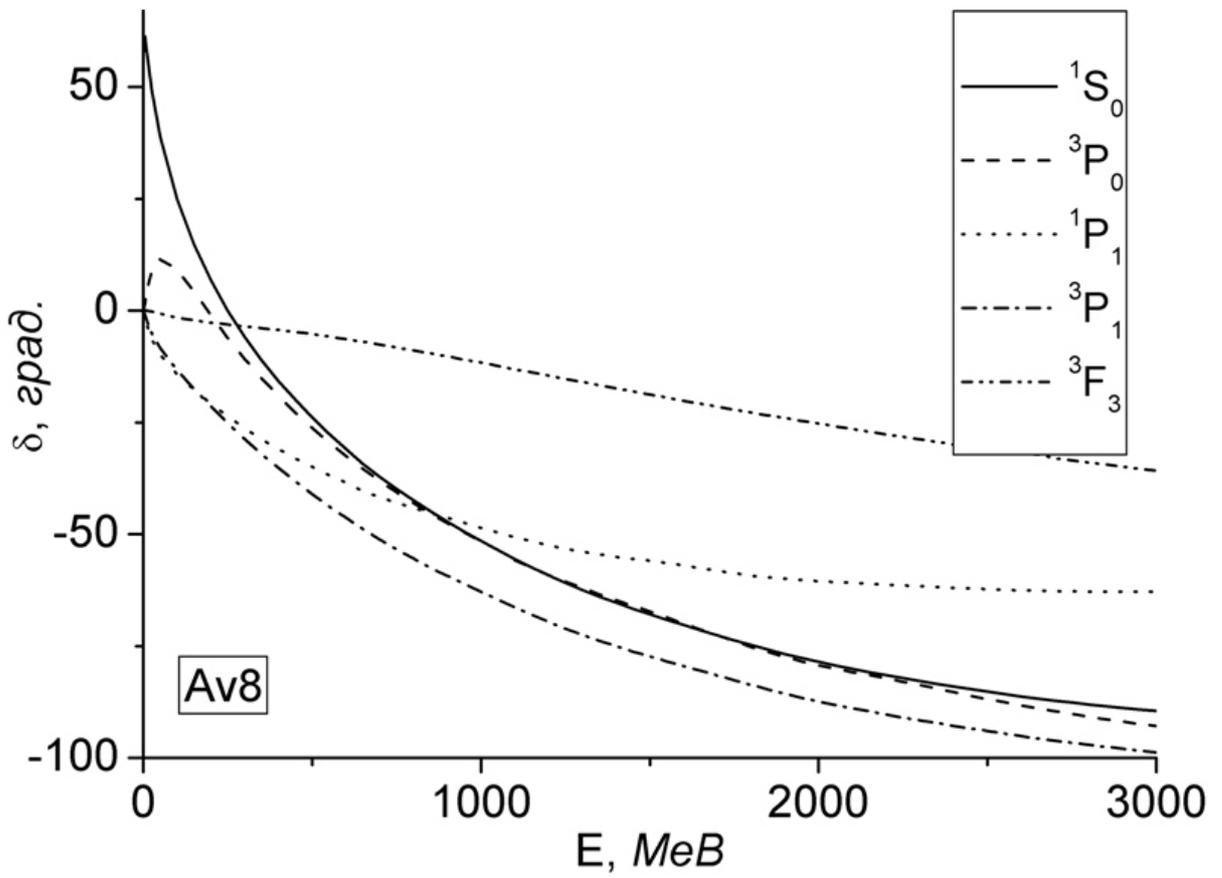

Рис. 7. Фазові зсуви *np*- розсіяння для потенціалу Av8'sscc

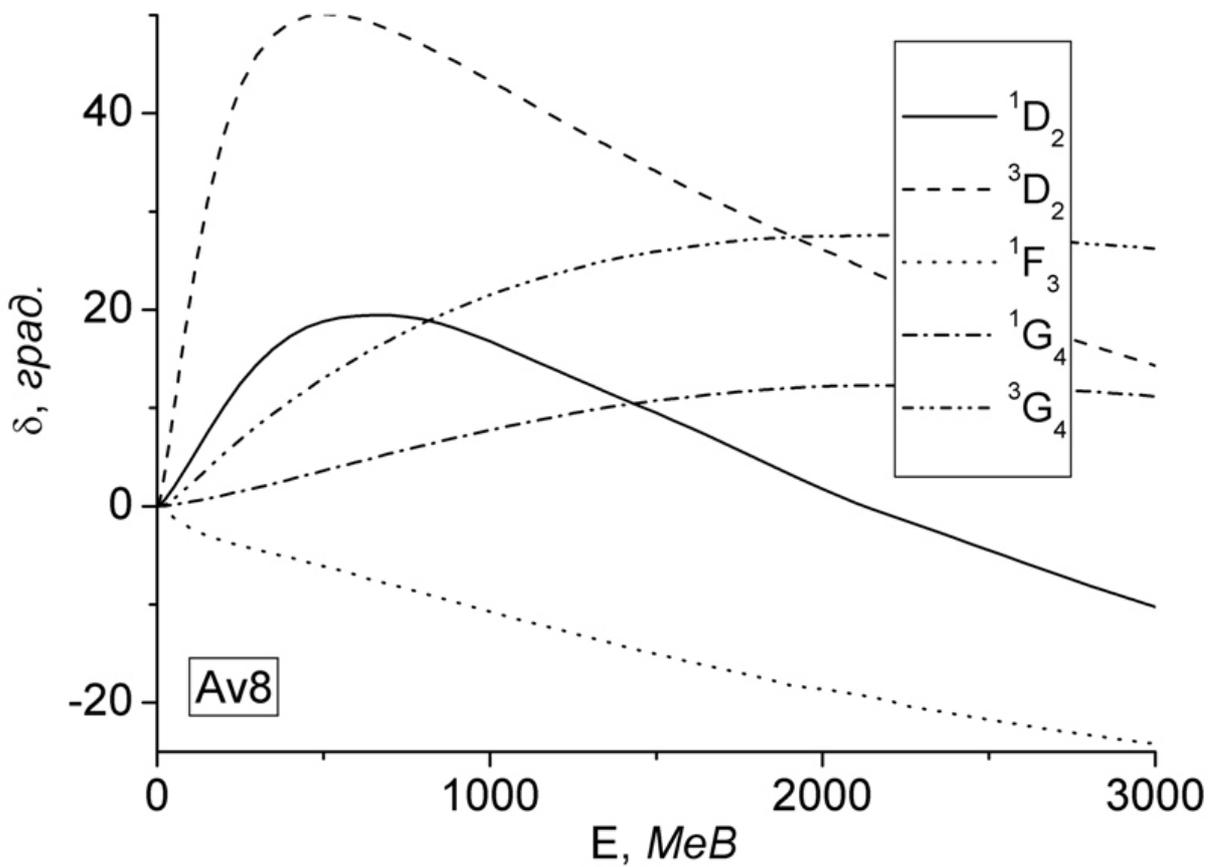

Рис. 8. Фазові зсуви *np*- розсіяння для потенціалу Av8'sscc

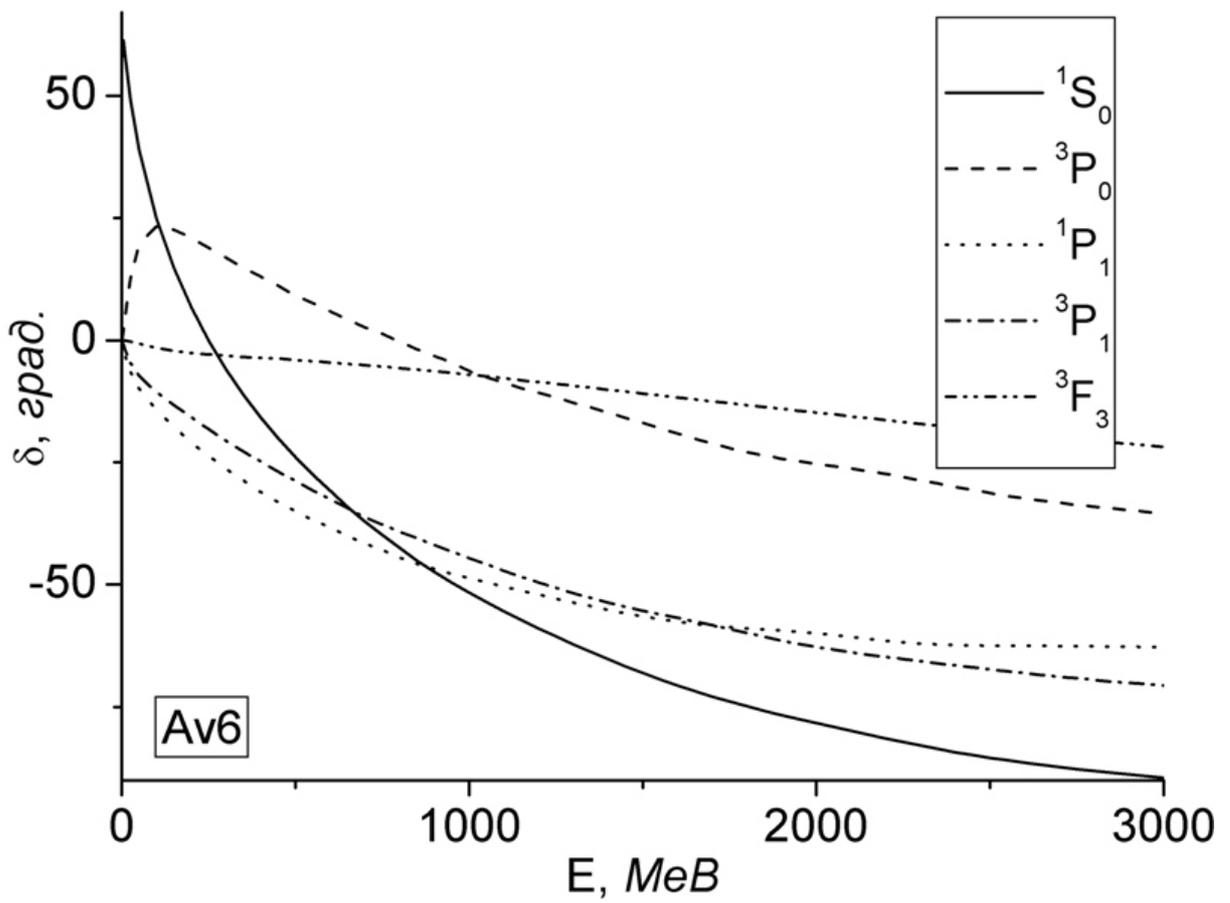

Рис. 9. Фазові зсуви *np*- розсіяння для потенціалу Av6'

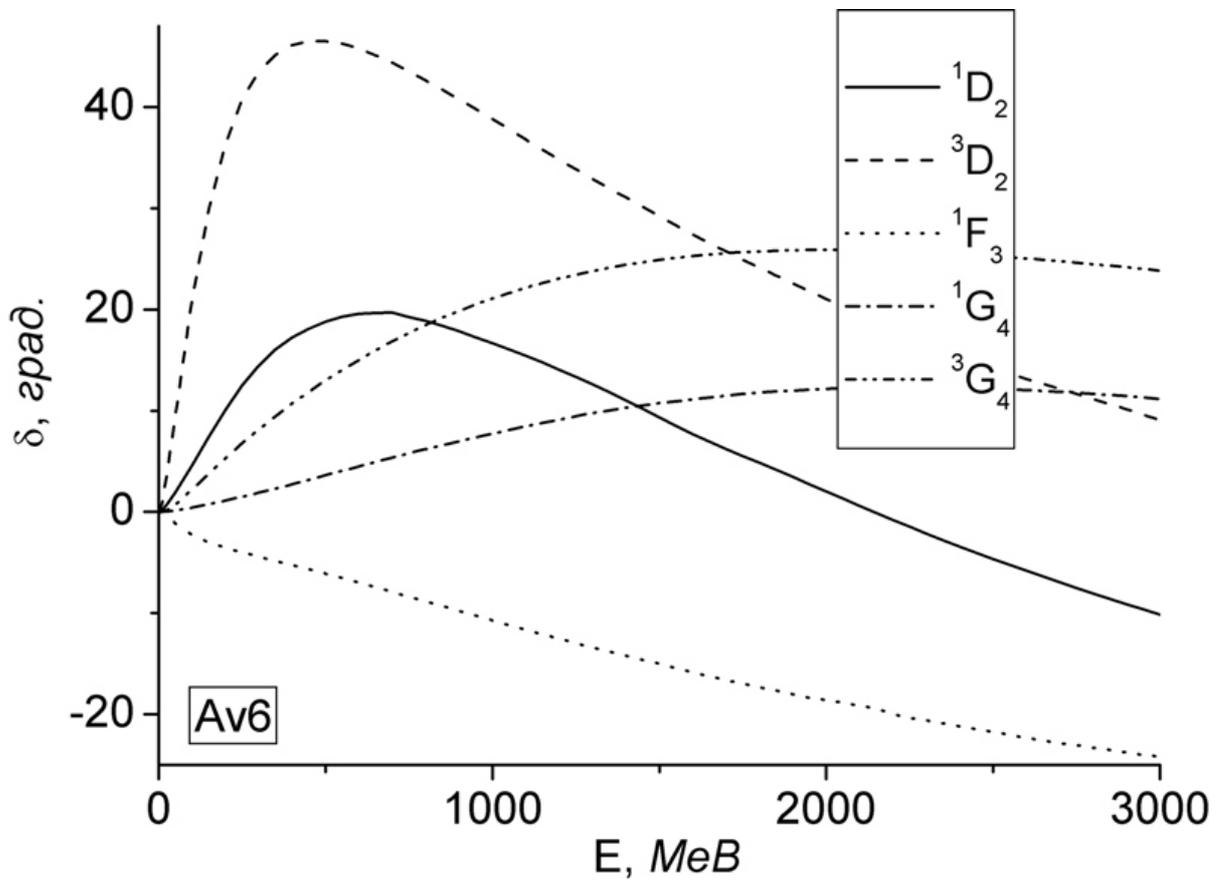

Рис. 10. Фазові зсуви *np*- розсіяння для потенціалу Av6'

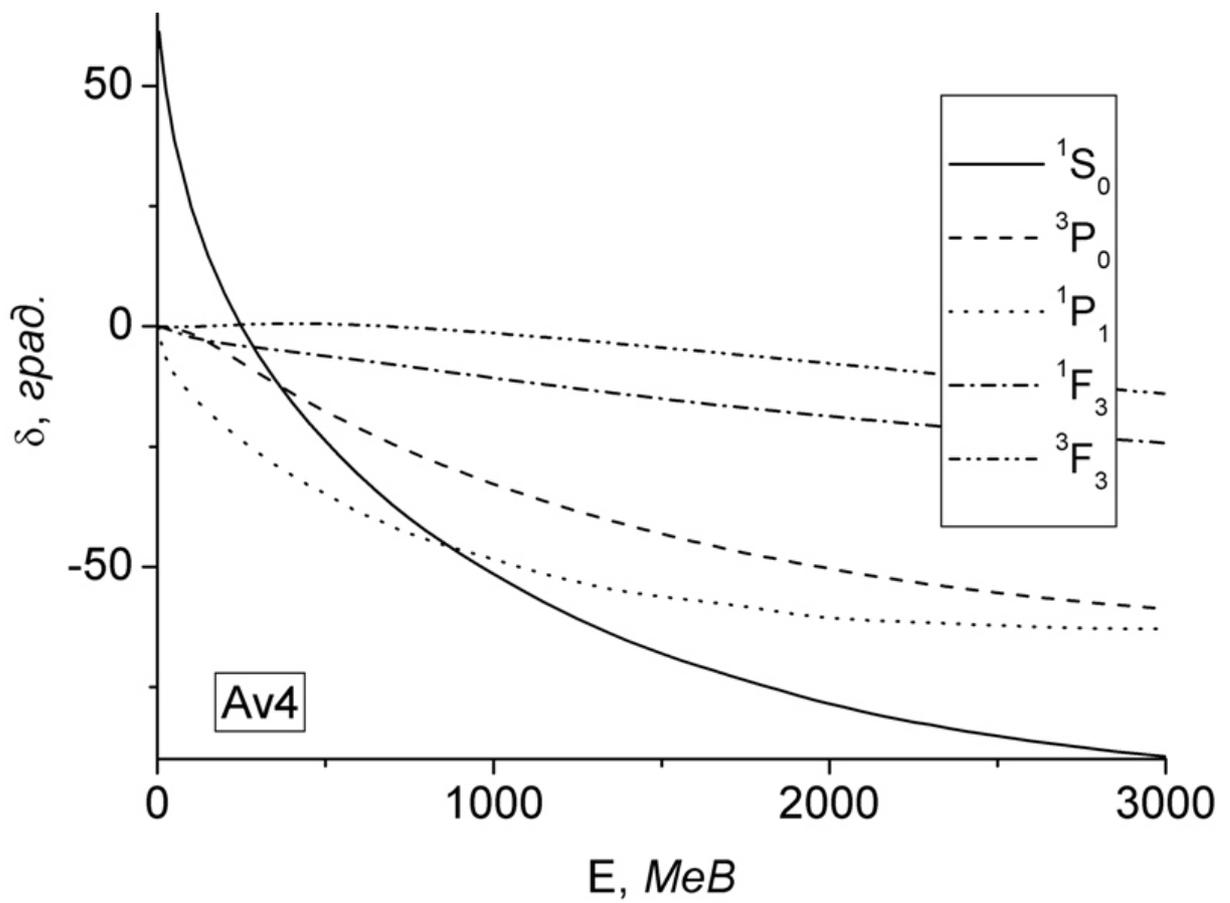

Рис. 11. Фазові зсуви *np*- розсіяння для потенціалу Av4'

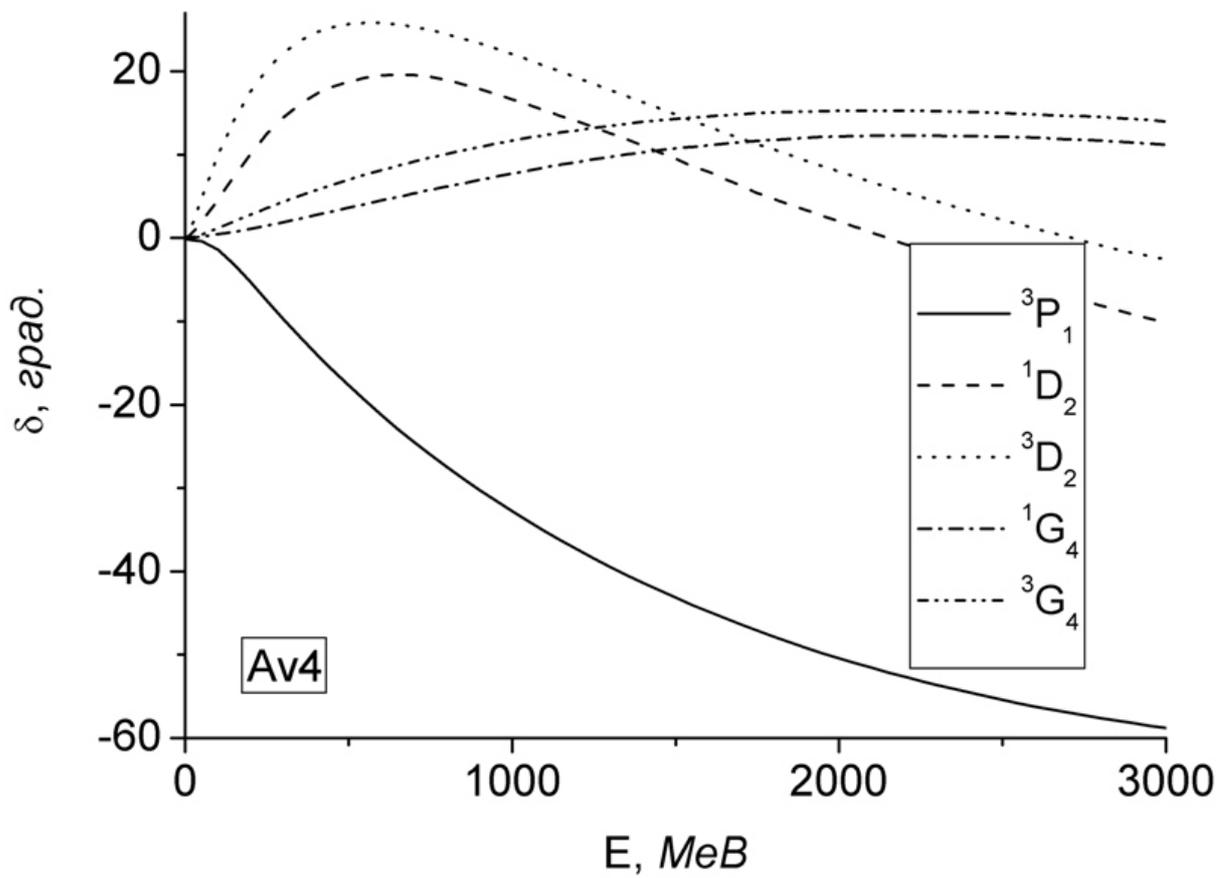

Рис. 12. Фазові зсуви *np*- розсіяння для потенціалу Av4'

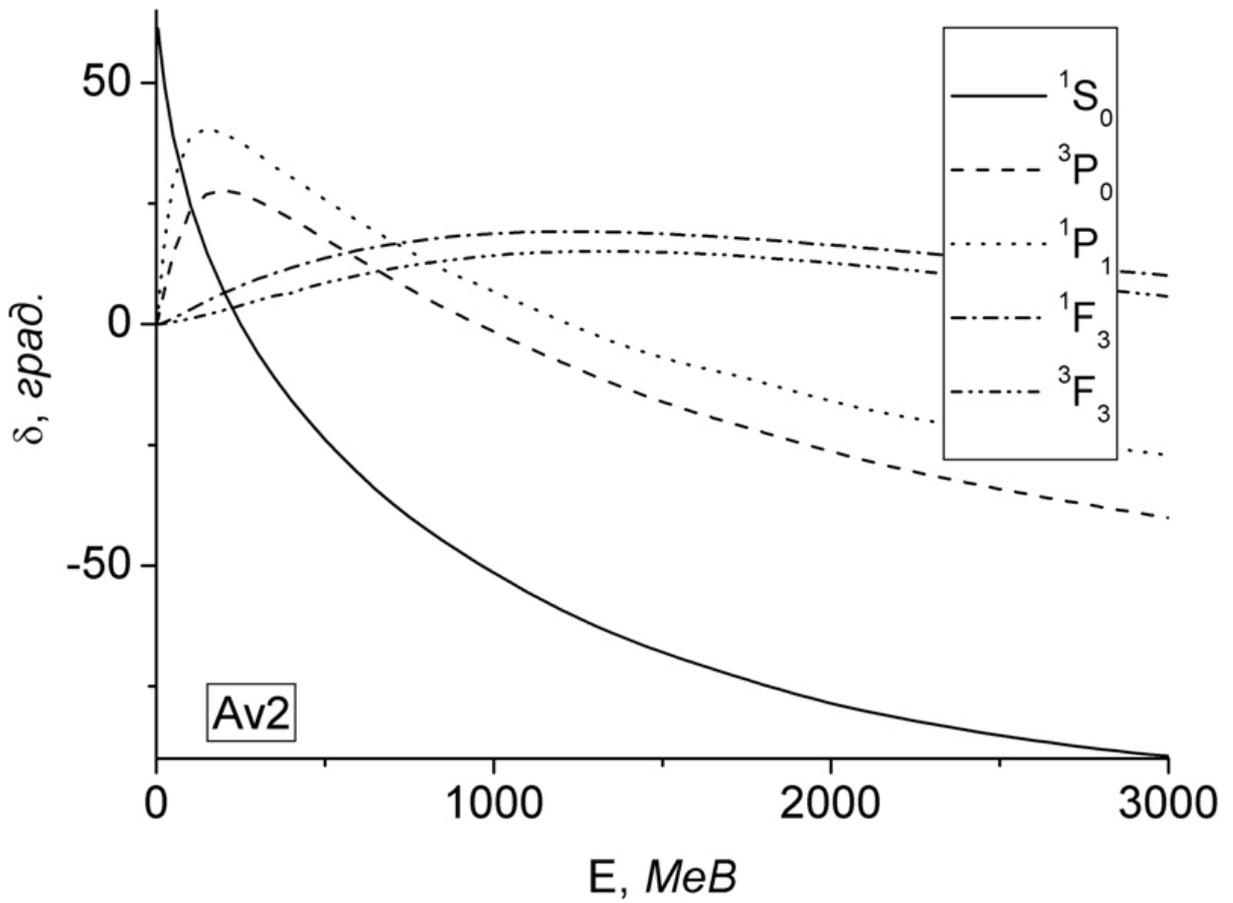

Рис. 13. Фазові зсуви *np*- розсіяння для потенціалу Av2'

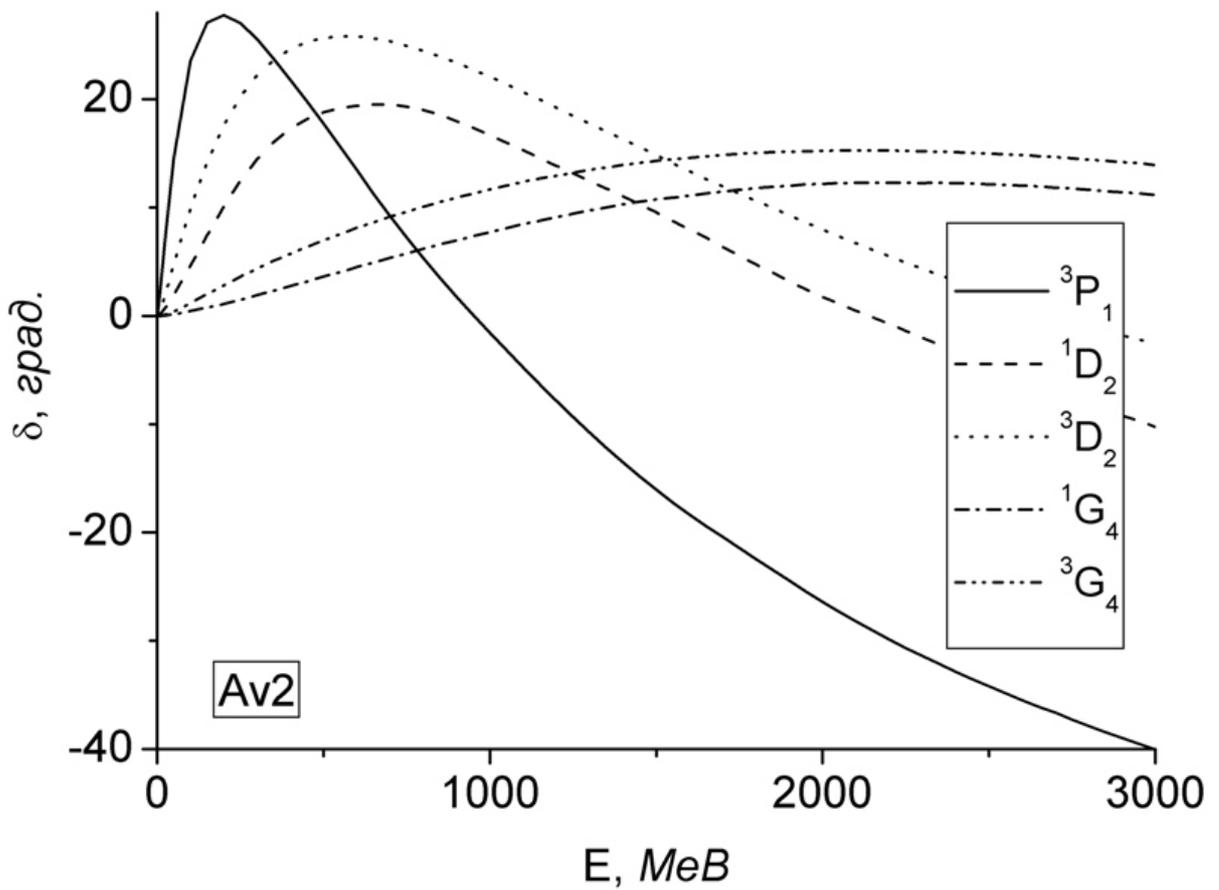

Рис. 14. Фазові зсуви *np*- розсіяння для потенціалу Av2'

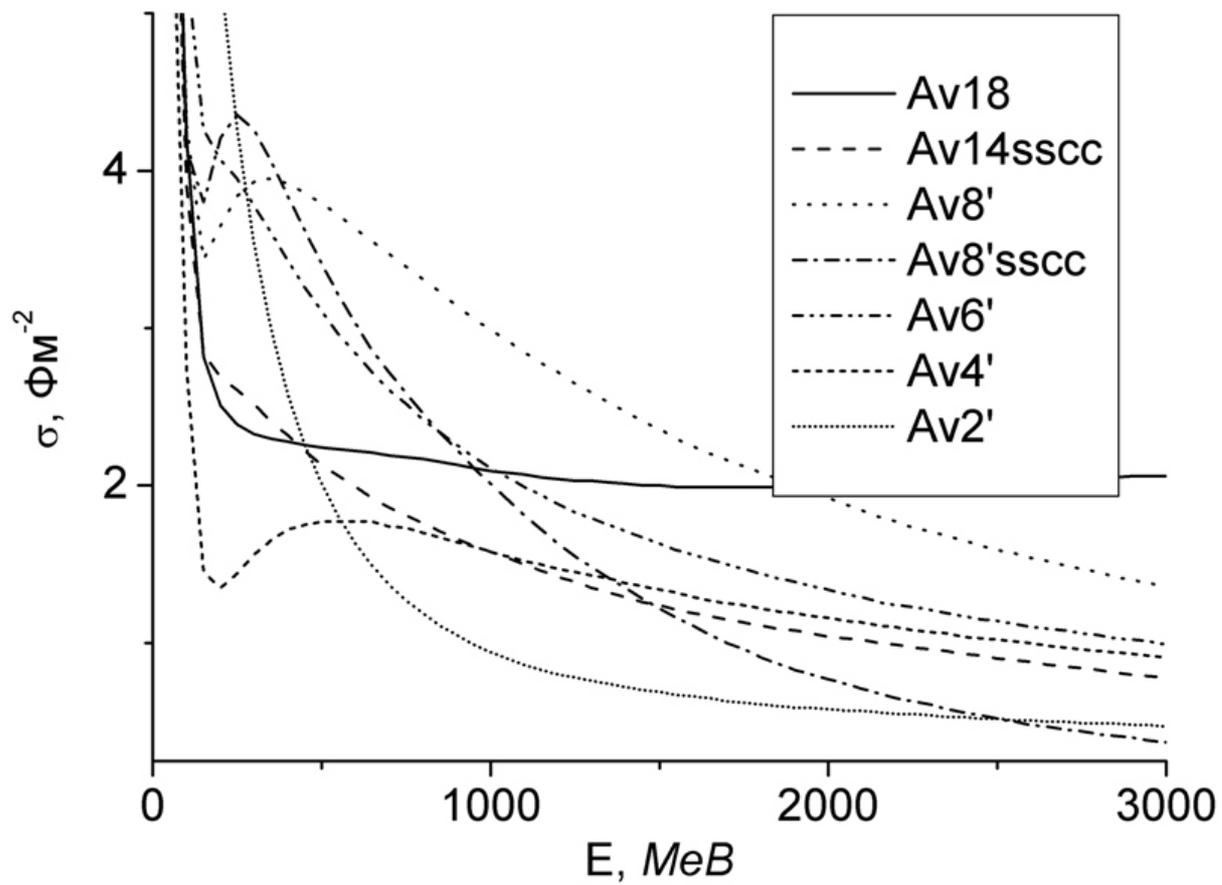

Рис. 15. Повний переріз *np*- розсіяння для потенціалів Аргоннської групи